**Lyman Continuum Emission Escaping from Luminous Green Pea Galaxies at z=0.5**

**Matthew A. Malkan**[1,2] **and Brian K. Malkan**[3]

**ABSTRACT**

Compact starburst galaxies are thought to include many or most of the galaxies from which substantial Lyman continuum emission can escape into the intergalactic medium. Li and Malkan (2018) used SDSS photometry to find a population of such starburst galaxies at z~0.5. They were discovered by their extremely strong [OIII]4959+5007 emission lines, which produce a clearly detectable excess brightness in the i bandpass, compared with surrounding filters. We therefore used the HST/COS spectrograph to observe two of the newly discovered i-band excess galaxies around their Lyman limits. One has strongly detected continuum below its Lyman limit, corresponding to a relative escape fraction of ionizing photons of 20+/-2%. The other, which is less compact in UV imaging, has a 2-sigma upper limit to its Lyman escape fraction of <5%. Before the UV spectroscopy, the existing data could hardly distinguish these two galaxies. Although a sample of two is hardly sufficient for statistical analysis, it shows the possibility that some fraction of these strong [OIII] emitters as a class have ionizing photons escaping. The differences might be determined by the luck of our particular viewing geometry. Obtaining the HST spectroscopy, revealed that the Lyman-continuum emitting galaxy differs in having no central absorption in its prominent Lyα emission line profile. The other target, with no escaping Lyman continuum, shows the more common double-peaked Lyα emission.

Key words: galaxies: ISM – galaxies: starburst – galaxies: dwarf – galaxies: stellar content – reionization, first stars – ultraviolet: galaxies

---

[1]   Corresponding author, email address: malkan@astro.ucla.edu

[2] Department of Physics and Astronomy, UCLA,  Los Angeles, CA 90095-1547, USA

[3] Crespi Carmelite High School, 5031 Alonzo Ave.,  Encino, CA 91316, USA

# 1. Introduction—Finding the Population of Starburst Galaxies that Reionized the Universe.

One of the most important events in the history of the Universe was the cosmic re-ionization. But our understanding of it remains limited, because we do not know where and when the ionizing photons originated. Although quasars are powerful enough sources to ionize local bubbles, they were too rare to have re-ionized the entire intergalactic medium (IGM)(Mitra et al. 2018; Parsa et al. 2018; Puchwein et al. 2019). The only remaining candidates which were sufficiently numerous are star-forming galaxies. Thus the question of re-ionization comes down to this: What ionizing radiation escapes from young star-forming galaxies?

Unfortunately, direct observations of ionizing photons escaping from galaxies are difficult, and when they have been done, they often yielded only upper limits. In particular, we first used HST's UV imaging sensitivity to search for Lyman limit continuum (LLC) leaking out of star-forming galaxies at z~1 (Malkan et al. 2003). Further FUV observations of large samples of emission-line galaxies have also set limits on the relative LLC escape fraction (compared to the expected intrinsic Lyman break strength) of <2% at z=1 (Rutkowski et al. 2016), <6% at z=2-2.5 (Rutkowski et al. 2017; Matthee et al. 2017), and <21% at z=3.08 (Siana et al. 2015; Japelj et al. 2017).

More recent works have begun to detect escaping LLC from galaxies at low redshifts (Izotov 2016a,b, 2018a,b) and at higher redshifts (deBarros et al. 2016, Bian et al. 2017, Vanzella et al. 2018, Rivera-Thorsen et al. 2019). A recent spectroscopic survey of 124 z~3 galaxies has detected 15 of them (weakly) below the Lyman limit, specifically those with the strongest Lyman-α emission lines (Steidel et al. 2018). Converting the measurements into estimates of the intrinsic escaping Lyman continuum fluxes at high redshifts is complicated by the uncertainty in the corrections for IGM absorption.

Observations at lower redshift do not suffer absorption by intergalactic HI. However, until recently most EUV observations of local starbursts have also failed to detect escaping LLC. FUSE obtained one possible detection, of the dwarf starburst Haro 11 at $f_{esc}$ = 3.3% (Leitet et al. 2011). HST/COS has detected a 21% escape fraction in a starburst galaxy at z=0.23 (Borthakur et al. (2014), and two other dwarf starbursts with relative escape fractions of 21% (Leitherer et al. 2016). Corrected for internal dust extinction, these correspond to absolute escape fractions of 1--8%.

Simply targeting blue galaxies has not proved efficient—the key is to identify the subset of galaxies which is particularly likely to release ionizing photons. Recent work suggests that the best candidates are the most extreme young starbursts, which concentrate their energy release in very compact regions (Izotov 2018a,b). In the local Universe at least, these characteristics tend to be found in metal-deficient dwarf galaxies. One of their defining properties is very strong [OIII]-emission line doublets, which gives them the name 'green peas' (Cardamone et al. 2009; Izotov et al. 2011; Nakajima and Ouchi 2014).

Extremely strong [OIII] emission (EW([OIII]) = 200--2000Å) and extremely high [OIII]/[OII] ratio >5 appear to be predictors of leaking LLC (Vanzella et al. 2020). Izotov et al. (2016a, 2018b) found that the latter is a necessary though not sufficient predictor, since high [OIII]/[OII] correlates with stronger LLC escape, with substantial scatter. At z=3.2 another such green pea (EW([OIII])~1000Å) has a recent spectroscopic detection of its LLC which indicates a relative escape fraction of 64% (de Barros et al. 2016). A recent HST imaging study of the Lyman limits of 61 galaxies at z=3.1 (Fletcher et al., 2019) finds that 12 of them have detectable escaping LLC photons, with $f_{esc}$ of 15% or higher. Four out of eight of these detected Lyman leakers have strong [OIII] lines with rest-frame 5007+4959 equivalent widths of 100--1000Å. And the [OIII]/[OII] ratios, where measured, were 6--10 (Nakajima and Ouchi 2014). In the midst of these recent discoveries, we used a new search for green peas at redshift z~0.5 that were bright enough for FUV spectroscopic followup with the Hubble Space Telescope.

## 2. New [OIII]-Strong Emitting Green Pea Galaxies at z~0.5

The original "green pea" galaxies were discovered at redshifts around z~0.2, which shifts their strong [OIII]5007+4949 emission line doublet into the r bandpass (Cardamone et al. 2009). We have recently again used broadband *photometry*--pushing SDSS imaging to its limits--to extend the search for green peas out to z=0.6 (Li and Malkan 2018; LM18). Around this redshift, the [OIII] doublet is shifted into the i band. And if its equivalent width is higher than $EW_{rest}>200$Å, it detectably brightens the entire broadband magnitude by several tenths of a magnitude relative to the flanking r and z photometric bands. There is of course a continuous distribution of [OIII] equivalent widths; we select this operational threshold because it is the starting point at which large low-resolution or photometric surveys work well.

LM18 had defined a photometric sample of 5169 extreme emission-line galaxies with strong i-band-excesses. These included contaminants, in which this excess is from strong Hα emission at z~0.15, and/or from an active galactic nucleus. Most of those are screened out by additional photometric cuts, including IR photometry from WISE. This narrowed the sample of photometric candidates down to 968. But spectroscopic confirmation was needed. With 7 nights of spectroscopy in 2018--19 with the Kast dual spectrograph on the Shane 3-m telescope at Lick Observatory, plus existing SDSS spectra, we have now confirmed that in a randomly selected sub-sample of 33 of these galaxies, the strong emission line in the i filter is, indeed, [OIII]5007+4959 from an extreme starburst, not an active galactic nucleus (Malkan and Chang 2020). Next, we discovered that these intermediate-redshift [OIII] emitters are so extremely blue, that they are easily detected in the GALEX survey, in both its NUV and FUV filters. This means that their rest-frame continuum down to the Lyman limit is bright enough for spectroscopy with HST.

## 3. Observational Data

### 3.1 Target Galaxy Selection

If a galaxy's redshift is too low, its crucial Lyman limit wavelengths (LLC: $\lambda_0 < 912$Å) lie in the dropping sensitivity of HST's UV-detecting instruments. If the redshift is high (z>2), the galaxies are extremely faint, and also suffer from an uncertain amount of intergalactic HI absorption. It turns out that a sensitive LLC search can be made at z~0.5. Although the higher redshift makes a given galaxy fainter than the original "green pea" galaxies, this is compensated for by the fact that the COS spectral sensitivity of the G140L configuration is near its peak at an observing wavelength of 1250Å[1].

We therefore selected from our new i-band excess sample the two green pea galaxies with the highest [OIII] luminosities, which are nonetheless representative of the entire class of z=0.5 green peas. Their higher redshifts (z=0.48, z=0.54) make J090334+174717 (denoted OIII-10 in LM18) and J020748+004735 (denoted OIII-17) easier targets than most previous low-redshift starbursts which have been searched for escaping LLC. For comparison, the detection of a relative LLC escape fraction of 21% in SDSS0921+4509 required 9 HST orbits (Borthakur et al. 2014), and comparable COS integrations were needed to weakly detect LLC in two of three other similar galaxies which were not green peas (Leitherer et al. 2016).

### 3.2 Previous Observations

#### 3.2.1 Spectroscopy

The rest-frame optical spectra of the two target starburst galaxies, from the Sloan Digital Sky Survey (SDSS, with spectrum ID's 0702-52178-416 for J020748+004735, and 5297-55913-0765 for J090334+174717, included in Figure 1), show absolutely no trace of any broad Hβ wings, [NeV] of any other indicator of Seyfert nuclear activity. J020748+004735 is mistakenly identified as a "QSO" in NED. That is a common Sloan misclassification because of the extremely blue broadband colors (Cardamone et al. 2009).

---

[1] https://hst-docs.stsci.edu/display/COSIHB/5.1+The+Capabilities+of+COS#id-5.1TheCapabilitiesofCOS-Section5.1.15.1.1First-OrderSensitivity

Table 1. Optical observed and derived properties of target green pea galaxies, from LM18. The last 3 columns are the parameters from SED model-fitting in LM18.

| Galaxy Name | R.A. 2000 [hh mm ss] | Dec. 2000 [dd mm ss] | $z$ | $D_L$ [Mpc] | $O_{32}^2$ ... | $EW_0$(OIII) [Å] | log[O/H] +12.0 | logM* [$M_\odot$] | Age [Myr] | SFR [$M_\odot$/yr] |
|---|---|---|---|---|---|---|---|---|---|---|
| OIII-17 (J0207) | 02 07 48.0 | +00 47 35 | 0.540 | 3190 | 7.4 | 455 | 8.33 | 9.73 | 87 | 56 |
| OIII-10 (J0903) | 09 03 34.6 | +17 47 18 | 0.478 | 2770 | 6.5 | 575 | 8.24 | 9.89 | 56 | 34 |

All emission line fluxes are taken from tables published in LM18, after re-checking that they are correct measurements from the SDSS spectra. Luminosity distances with the standard cosmology and given in the fifth column. Our two prime target galaxies listed in Table 1, named OIII-10 And OIII-17 in LM18, have a very highly ionized ISM, with $O_{32}$ = [OIII]5007+4959/[OII]3727= 7.4 and 6.5, and [OIII]5007+4959/H$\beta$ > 7. Figure 1 shows their BOSS spectra. The rest-frame equivalent widths of their [OIII]5007 lines are 575 and 455Å, comparable to what is observed in typical green peas at lower redshifts. Not surprisingly, these targets are minimally reddened; their Balmer decrements give E(B-V) of 0.16 and 0.18 mags. Independently, the strong He I emission lines in J090334+174717 allow us to measure the reddening-sensitive line flux ratio of the triplet lines $\lambda\lambda$5876/3187Å to be 3.9+/-0.6. This can be compared with the predicted intrinsic recombination ratio of 2.7 (Osterbrock 1989, Table 4.6). For a standard interstellar reddening law, this ratio agrees with our estimate from the Balmer decrement of E(B-V)=0.16+/-0.05 mag. Nearly all the reddening is internal, based on the very small Milky Way extinctions estimated by the NASA Extragalactic Database[3]--E(B-V)=0.02 for both galaxies.

---

[2] F([OIII]5007+4959)/F([OII]3729+3726)
[3] http://ned.ipac.caltech.edu/

The weak but crucial auroral [OIII]4363 emission line is well detected in both optical spectra. This provides a direct measure of the electron temperature, which then provides a 'gold standard' estimate of the [O/H] abundance (Ly et al. 2014, 2016). Our two target galaxies have O/H abundances roughly a third of solar, and N/H is much more sub-solar. Finally, the redshifts of these targets are high enough to show--with SDSS optical spectroscopy--another characteristic associated with strong LLC leakage: clearly detectable MgII 2800Å emission (Henry et al. 2018).

### 3.2.2 Photometry.

An extensive range of additional observations are available to provide further diagnostics of these two galaxies. Table 2 presents broadband photometry in the UV, taken from GALEX[4], in the optical, taken from the Sloan Digital Sky Survey[5], and in the infrared from WISE[6]. The broadband spectral energy distributions of both targets are included as solid squares plotted in Figure 1. The second-to-reddest SDSS broadband filter is the i-band, which is enhanced relative to the surrounding r and z filters by the extremely strong [OIII]5007+4959 emission doublet flux. One of the targets, J090334+174717, also has SDSS spectroscopy extending into the near-infrared, providing measurements of additional key red emission lines.

Table 2. Broadband photometry of target green pea galaxies, with fluxes given in microJanskys for entire galaxy

| Fluxes | SDSS | | | | | UKIDDS | UKIDDS | GALEX | | WISE | | |
|---|---|---|---|---|---|---|---|---|---|---|---|---|
| | *u* | *g* | *r* | *i* | *z* | *Y* | *J* | FUV | NUV | *W1* | *W2* | *W3* |
| NAME | %err | %err | %err | %err | %err | %err | %err | %err | %err | %err | %err | %err |
| OIII-10 =J0903 | 32 (5) | 35 (2) | 44 (2) | 88 (2) | 64 (6) | --- | --- | 12 (34) | 29 (15) | 84 (08) | 78 (17) | 714 (27) |
| OIII-17 =J0207 | 40 (4) | 37 (2) | 46 (3) | 72 (3) | 42 (8) | 77 (13) | 46 (26) | 18 (15) | 32 (15) | 36 (14) | 44 (25) | <470 |

---

[4] http://galex.stsci.edu/GalexView/
[5] http://skyserver.sdss.org/dr15/en/tools
[6] https://irsa.ipac.caltech.edu/cgi-bin/Gator/

Both of our target galaxies have repeated solid GALEX detections in both the NUV and FUV channels. The observed NUV fluxes (bandpass runs from 1771—2831Å, with effective wavelength of 2270Å) were independently measured from multiple GALEX observations, entirely longward of the Lyman limits of these galaxies. The rest-frame 1450Å continuum fluxes measured by GALEX from PSF-fitting (neglecting any extended flux) are 10 and 32 µJy for the two target galaxies. We include WISE detections in bands W1, W2 and W3 (3.4, 4.6 and 12 microns), given in Table 2. The relatively red W1 – W2 colors of J090334+174717 and, especially, of J020748+004735 are produced by hot dust, as is seen in other extreme starburst galaxies with high equivalent widths of Hβ (Izotov et al. 2011).

Compared with the dwarf galaxies which have previously detected Lyman limit continuum leakage, the stellar masses of both of our target galaxies ($\sim 7*10^9$ M$_\odot$; LM18) are at the upper end of the distribution (Table 1). Their high rates of current star formation ($\sim 40$ M$_\odot$/year) are also exceptionally high, a factor of several times those of the most extreme green peas found at z~0.1 (Izotov et al. 2011; Jaskot and Oey 2013). Both of our targets have [OIII] luminosities exceeding $10^{43}$ erg sec$^{-1}$, several times larger than the most luminous previously studied "green pea" galaxies at lower redshifts (LM18).

### 3.3 HST/COS Observations

#### 3.3.1 Ultraviolet Spectroscopy

In the Supplemental round of Cycle 26, we obtained 3 HST Guest Observing time to observe our two target galaxies with COS, for 3 orbits each. The spectra of the two galaxies were obtained in program GO-15471, in 2018 November and December[7]. First, short acquisition NUV imaging observations of one minute were obtained using the MIRROR A option (see Figure 5, in Section 3.3.5). Then the remainder of the first orbit, and all of the remaining two orbits were devoted to long spectroscopic integrations with the G140L grating, through the PSA aperture, with the galaxy at Lifetime Position 4. The total spectroscopic integration times were about 8200 seconds per target, taken in ACCUM mode. The reduced, extracted 1-D spectra of the entire source were then binned combining groups of 7 pixels, giving 0.54Å final pixels, small enough to preserve the instrumental R~2000 resolution. The starlight continuum of each galaxy is very well detected longwards of their Lyman limits, including several strong absorption lines.

---

[7] http://www.stsci.edu/cgi-bin/get-proposal-info?id=15471&observatory=HST

### 3.3.2 Overall Spectral Energy Distributions

Figure 1 presents the combined multiwavelength spectral energy distributions (SEDs) of the two COS target galaxies. The GALEX fluxes (solid blue squares) are consistent with the observed continuum fluxes measured in our COS spectra (shown by the black lines). The SDSS ugriz photometry (solid blue squares) is also consistent with the SDSS spectra (shown by the red lines).

These SEDs were already fitted by LM18, with double stellar population models (an old population dominating the mass, combined with a recent burst of star formation). Their SED model-fitting parameters are given in the last 3 columns of Table 1. The UV continua of both galaxies are dominated by large numbers of very recently formed massive stars, as expected from their extremely blue colors and strong line emission.

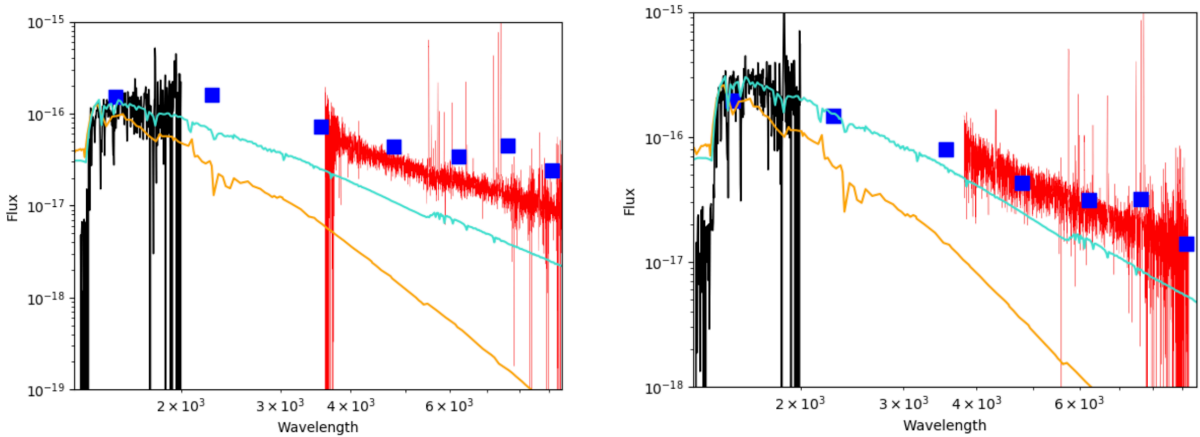

Figure 1: Observed UV/optical spectra of the two target galaxies: J090334+174717 (OIII-10) on the left and J020748+004735 (OIII-17) on the right. The black lines indicate the COS spectra (see Figure 3 for a zoomed-in view around the Lyman limit spectral region). The red lines show the observed SDSS spectra. The solid blue squares represent the broad band photometry from GALEX and SDSS listed in Table 2. The orange curve shows a redshifted,

reddened stellar atmosphere model spectrum for an O star ($T_{eff}$=30,000K) from Hainich et al. (2019). The turquoise curve shows the young stellar population component taken from the model fits of LM18. Note that the vertical flux scale of the left graph extends an additional order of magnitude fainter, to show the faint upper limits to the (undetected) Lyman continuum of J090334+174717.

### 3.3.3 Lyman Limit Observations

We examined both the original individual 2-D and extracted, calibrated 1-D spectra for both target galaxies. In the full COS spectrum of J090334+174717 no flux is detected below the Lyman limit. However, significant flux is clearly detected beyond the Lyman limit of J020748+004735 (i.e. at observed wavelengths shorter than 1406Å). Before identifying this as genuine Lyman continuum leaking from the galaxy, it is imperative to rule out all other possible instrumental sources of spurious weak continuum signals that could artificially produce such an apparent flux. We already know that none of these problems occurred in the otherwise identical COS spectroscopic observation of J090334+174717.

We have accordingly made several tests of the reality of the Lyman limit continuum in J020748+004735. In particular, we were concerned that dark counts, or other background light, including scattered Earthlight, might possibly have created spurious far-UV ($\lambda_0 < 1406$Å) continuum in the 1-D spectrum. In the first test, we modified the pulse-height threshold acceptance criteria for counting genuine photons. We did this by reprocessing the raw 2-D corrtag_a.fits photon files, modifying the values of PHALOWRA and PHAUPPRA away from their default values of 2 and 23, respectively. None of these modifications changed the strength of the continuum detected at any wavelengths including at $\lambda_0 < 1406$Å. Next we considered the possible contamination from scattered Earthlight, although by far the strongest signal is within 5Å of the 1304Å O I geocoronal emission line--far from the wavelengths of interest. We separated several sub-spectra based on the solar angle above or below the Earth's terminator during the observations. The strength of O I geocoronal emission varied enormously from the observations obtained in shade, to those obtained when the sunlit face of the Earth was visible. But again, in none of these different spectra was there any difference in the strength of the continuum, including the detected far-UV $\lambda_0 < 1406$Å. Finally we analyzed the raw 2-D spectrum of individually detected photons above and below the Lyman limit, as shown in Figure 2. The narrow solid rectangle shows the Lyman continuum extraction region, which consists of 17 spatial rows. Inside this region, from columns corresponding to wavelengths of 1375 to 1406Å (rest wavelengths of 892--912Å, all beyond the Lyman limit), there are a total of 344 detected photons. The total rate of background and dark current photons in this wavelength range was determined from the 30--50 pixel wide rectangles above and below the galaxy spectrum, shown with the dashed line rectangles. Both regions indicate that the total number of background/dark photons within the galaxy spectrum box should be 81. Since the uncertainty in the detection of 263 net galaxy photons is dominated by Poisson statistics, the significance of the ionizing continuum is 14.2 times the uncertainty. This 14--1 SNR value was insensitive to changes in the sizes of the source and background extraction windows.

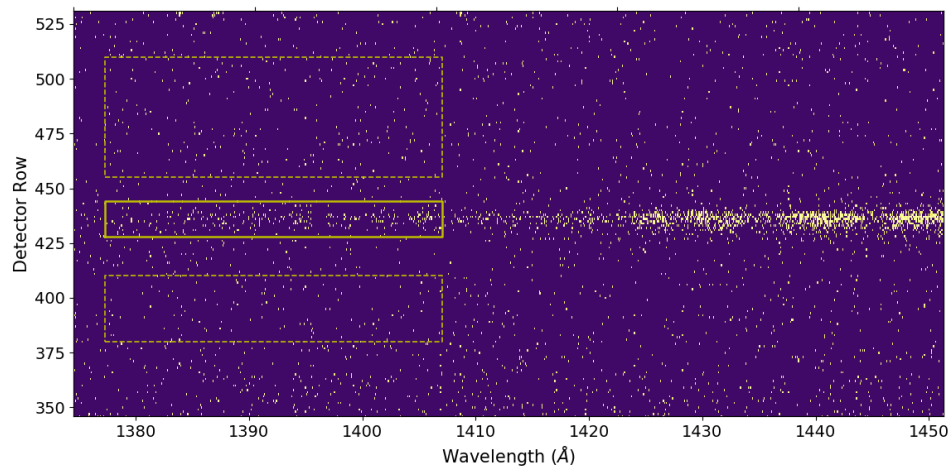

Figure 2. Blowup of the Far-UV portion of our two-dimensional COS spectrum of J020748+004735, showing all individual photons detected in the complete 8200 seconds of exposure. The solid box shows the extracted region of the galaxy spectrum below the Lyman limit. The two dashed boxes show the comparison regions where the background photon counts were determined. Compared with the background regions, the very significantly higher density of FUV photons detected in the rows containing the galaxy light is clearly evident to the eye.

### 3.3.4 Relative and Absolute Escape Fractions of Ionizing Photons

To determine the relative Lyman continuum escape fraction requires knowing the intrinsic flux drop from the red side to the blue side of the Lyman limit. The Lyman limit relative escape fraction can be defined as:

$$f_{esc,rel} = \{I(900)/I(940)\}_{obs}/\{I(900)/I(940)\}_{model}$$

where the intrinsic drop at the Lyman continuum is the measured ratio just below the Lyman limit to the flux just longward of rest wavelength 912Å, avoiding absorption lines. (Over such a small wavelength interval, the possible effect of dust reddening is negligible.) This observed quantity is compared with $\{I(900)/I(940)\}_{model}$, which is predicted from using stellar atmosphere models to extrapolate the observations at longer wavelengths. This is straightforward to estimate, since the EUV continuum is predominantly produced by the hottest (youngest) massive stars present, and these have similar Lyman limit jumps, according to stellar atmosphere models for a wide range of stars on the upper main sequence. For simplicity, we match the EUV continuum in both galaxies with a recent O star model from Hainich et al. (2019) adopting an effective temperature of 30,000 K and a surface gravity of log g = 3.4. We then reddened this model by the E(B-V) estimates obtained from the Balmer emission line ratios (Table 1, from LM18). The scaled reddened O star spectrum (orange line) is overplotted with the HST/COS + SDSS optical spectroscopy, and broadband photometry, in Figure 1.

This realistic model stellar atmosphere predicts that the intrinsic flux beyond the Lyman limit (900Å rest wavelength) is 0.3 times that of the 940Å continuum, and this is the standard assumed ratio in other Lyman limit studies (e.g., Reddy et al. 2016). Similar estimates would also have been obtained using almost any model of a population of recently formed stars. For example, LM18 fitted young stellar populations (age~70 Myr) to the optical spectra of both of these galaxies. As shown in Figure 1, these models predict a 900Å flux that is 0.28 times the 940Å continuum.

In J020748+004735, the average flux above the Lyman limit (rest wavelength 940Å-946Å) is $2.6(+/-0.2) \times 10^{-16}$ ergs sec$^{-1}$cm$^{-2}$Å$^{-1}$ (horizontal red line in Figure 3). The average flux below the Lyman limit (880Å-912Å) is $1.6(+/-0.1) \times 10^{-17}$ ergs sec$^{-1}$cm$^{-2}$Å$^{-1}$, summarized in Table 3. This gives a relative escape fraction, $f_{esc,rel} = 20(+/-2)\%$. In J090334+174717 the fluxes above and below the Lyman limit are $9.8 \times 10^{-17}$ and $< 2.2 \times 10^{-18}$, respectively. This sets a two-sigma upper limit of $f_{esc,rel} < 5\%$.

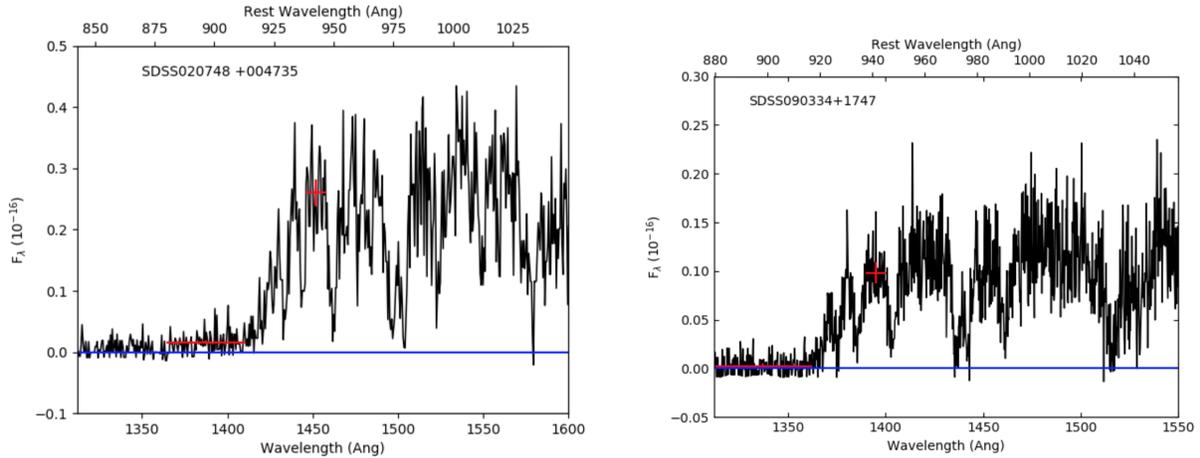

Figure 3: Zoom-ins of the observed COS spectra around the Lyman break region. Observed wavelengths are along the bottom x-axis. The rest-frame wavelength scales are shown on the top x-axis. In addition to OI 989Å and the OVI 1035Å doublet, about five absorption lines from the Lyman series are evident before the higher lines become blended. The horizontal blue lines show zero flux. Short red lines show the average of the continuum just longward of the Lyman break (940Å-946Å). Longer red lines show the average continuum shortward of the Lyman break (880Å-912Å). We do not extend the integration to shorter wavelengths because the rapidly dropping short-wavelength sensitivity increases the noise. Vertical red error bars show the one-sigma uncertainties in the integrated continuum fluxes in each bin. Note that the error bars for the fluxes beyond the Lyman limit--since they are averages of many independent pixels--are so small that they are only slightly larger than the thickness of the horizontal red lines. In J090334+174717 (right spectrum)--unlike J020748+004735 (left spectrum)--this ionizing flux is consistent with zero.

The relative Lyman escape fractions we have measured are sometimes called "dust-free" escape fractions, because they do not account for reddening within the observed galaxy or in the Milky Way (Borthakur et al. 2014). Making further assumptions, we can estimate the "absolute escape fraction" of ionizing photons that leave the galaxy without being absorbed by its interstellar dust. Following previous researchers, we define this as:
$f_{esc,abs} = f_{esc,rel} \times 10.**\{-0.4 \times A_{912,int}\}$.
Including this extinction requires extrapolating our optical reddening estimates down to the Lyman limit. Depending on the EUV reddening law assumed (Reddy et al. 2016), we estimate internal dust absorptions of $A_{912,int}$ = 1.5 -- 2.0 magnitudes in both galaxies. These imply absolute escape fractions of ~4% in J020748+004735, and <1% in J090334+174717. (Alternately, other authors normalize the Lyman limit continuum by the intrinsic (de-reddened) flux at 1500Å instead of 912Å (Izotov et al 2016a). We do not have spectroscopic observations at 1500Å, but we can roughly estimate that adopting the definition given in Equation 2 of Izotov et al 2016a would imply absolute escape fractions of ~7% in J020748+004735, and <1.5% in J090334+174717). Given the uncertainties about their internal EUV extinctions, we caution that these "absolute escape fraction" estimates could be in error by a factor of 2.

### 3.3.5 Lyman-α Emission lines

The COS spectra simultaneously cover the LLC, the FUV continuum redwards of the Lyman limit, and the complete Lyα line profiles. This removes uncertainties arising from calibration issues with separate observations, spatially extended emission, etc. Figure 4 shows zoomed in views of the Lyα emission line profiles from the two COS targets. The profile in J090334+174717 (right) has a central zero-velocity absorption, making it double-peaked. This profile is common in green pea galaxies (Henry et al. 2015), except that the blue peak is *stronger* than the red peak. There is no simple analytic form for these complex profiles; we fitted Gaussians merely to obtain a rough objective idea of the velocity spreads. J020748+004735 (left) is unusual in having no central dip, but instead a zero-velocity emission peak. If we attempt to characterize these complex profiles by a blue/red component separation velocity, as in Izotov et al. 2018b, it would be roughly 350 km/sec for J090334+174717, and <150 km/sec for J020748+004735.

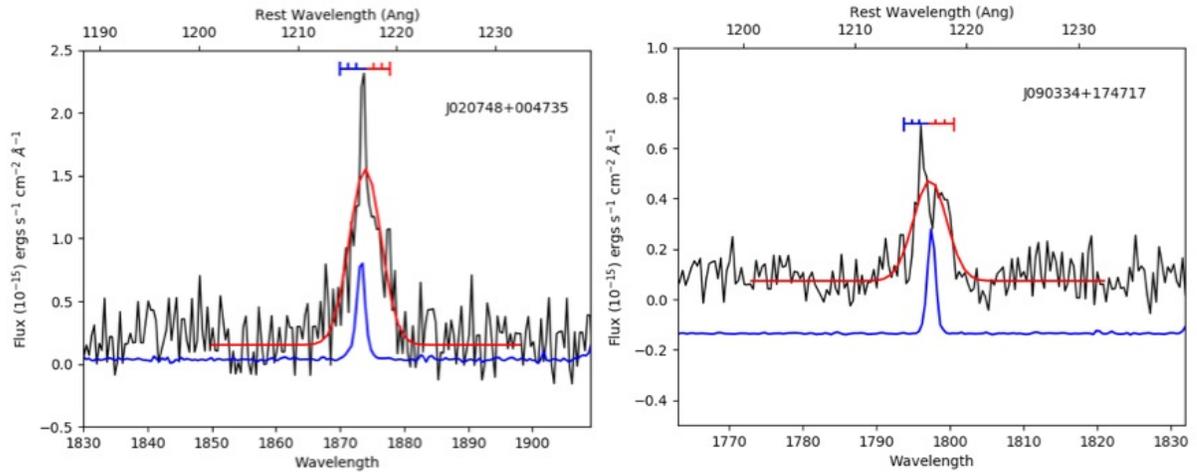

Figure 4. (Left): The black line shows the COS spectrum of J020748+004735 zoomed in around the Lyα emission line. The vertical scale shows the fluxes in units of $10^{-15}$ ergs sec$^{-1}$cm$^{-2}$Å$^{-1}$. The smooth red curve is the best fitting Gaussian which has a rest-frame equivalent width of −37 Å and a full width half max of 5.8 Å (900 km sec$^{-1}$). The red and blue horizontal lines extend to + and −600km sec$^{-1}$ from the rest velocity of the galaxy, encompassing the entire broad wings of the Lyα emission lines. (Thus the zero-velocity wavelength, based on the optical redshift, is at the boundary between the red and blue lines). Small tick marks show intervals of 200km sec$^{-1}$. The blue line shows the Hβ emission line profile over-plotted on the same velocity scale. (Right): Same as left panel, for the galaxy J090334+174717. This best fitting Gaussian in red has a rest-frame equivalent width equal to −21 Å and a full width half max of 5.3 Å.

Table 3: Observed HST/COS fluxes

| HST/COS Fluxes | Lyα Line flux | Continuum Flux around the Lyman Limit-Obsvd (ergs sec$^{-1}$cm$^{-2}$Å$^{-1}$) | | Continuum Flux around Lyman Limit-Dereddend (ergs sec$^{-1}$cm$^{-2}$Å$^{-1}$) | |
|---|---|---|---|---|---|
| GALAXY NAME | erg sec$^{-1}$ cm$^{-2}$ err | $f^+$ err | $f^-$ err | $f^+$ err | $f^-$ err |
| OIII-10 =J0903 | $2.2*10^{-15}$ $(0.3*10^{-15})$ | $9.8*10^{-17}$ $(2*10^{-17})$ | $<2.2*10^{-18}$ | $1.2*10^{-16}$ $(2.4*10^{-17})$ | $<2.6*10^{-18}$ |
| OIII-17 =J0207 | $8.6*10^{-15}$ $(0.7*10^{-15})$ | $2.6*10^{-16}$ $(0.2*10^{-16})$ | $1.6*10^{-17}$ $(1.1*10^{-18})$ | $3.1*10^{-16}$ $(0.2*10^{-16})$ | $1.9*10^{-17}$ $(1.3*10^{-18})$ |

$f^+$ is the flux above the Lyman Limit, between rest wavelengths of 940-946Å. $f^-$ is the flux below the Lyman Limit, between rest wavelengths of 880-912Å. In the last two columns, these fluxes have de-reddened for Milky Way extinction.

The Hβ emission line profile observed by SDSS is over-plotted on the same velocity scale. As expected for a small galaxy, it is spectrally unresolved, implying an intrinsic velocity FWHM of less than 150 km sec$^{-1}$. This is similar to the narrow core of the Lyα line. However, Lyα has much broader wings (FWHM=900 km sec$^{-1}$) which are presumably due to scattering in this optically thick resonance line. Comparing the observed Lyα fluxes (of 8.6 x 10$^{-15}$ and 2.2 x 10$^{-15}$ erg sec$^{-1}$ cm$^{-2}$) to the what is intrinsically predicted from the de-reddened Balmer line fluxes, assuming Case B recombination, gives Lyα escape fractions (Henry et al 2015) of 2.5% and 11% in J090334+174717 and J020748+004735, respectively. As will be discussed below, theoretical modelling has in fact predicted that larger Lyα escape fractions, and more symmetric centrally peaked Lyα emission line profiles, should be associated with higher Lyman continuum escape fractions (e.g. Dijkstra et al. 2016).

### 3.3.6 Spatial Distribution of UV Continuum from COS Imaging

Both target galaxies were spatially unresolved by GALEX, with its coarse angular resolution of 5 arcsec (FWHM). The SDSS images of J020748+004735 are virtually pointlike. J090334+174717 shows some faint extensions, which might indicate an interaction with a neighboring galaxy within a few Kpc. The features are confirmed in short-exposure NUV acquisition images from our HST visits which capture the bright central regions of the galaxies (see Figure 5). In particular, the UV surface density, and therefore the star formation surface density, is higher in J020748+004735. For both galaxies, all the detected UV flux was captured in the 2.5 arcsecond PSA in which the spectroscopy was obtained.

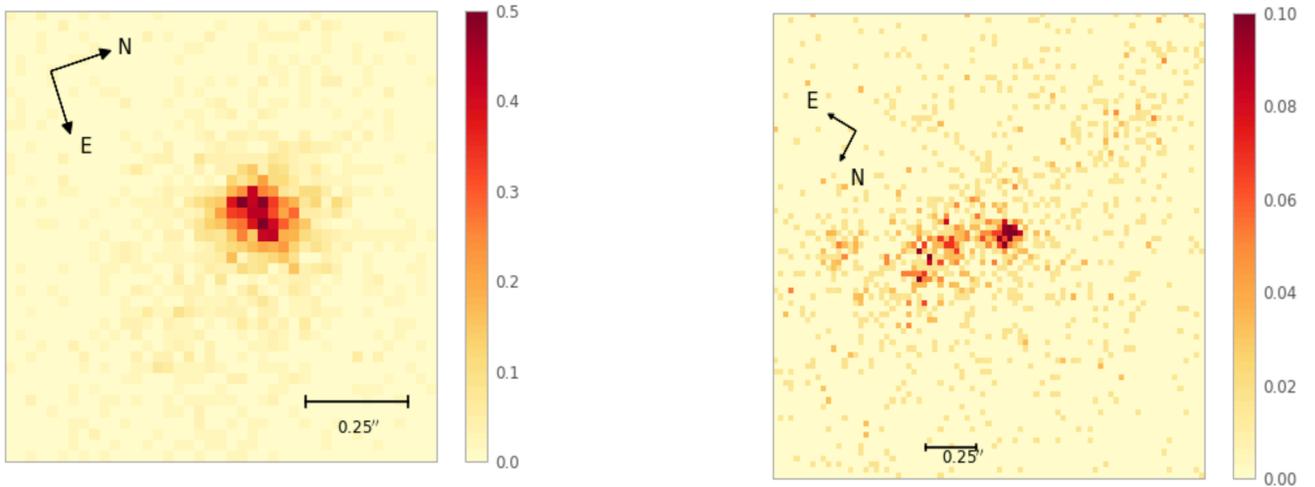

Figure 5: (Left) COS Direct NUV image of OIII-17 = SDSS J020748.0+004735. The entire field is one arc second per side. The horizontal bar shows a 0.25 arcseconds scale, which corresponds to approximately 1.6 Kpc at the redshift of these galaxies. (Right) COS Direct NUV image of OIII-10 = SDSS J090334.6+174718. The entire field is two arc seconds per side.

The SDSS spectroscopic survey for Luminous Red Galaxies happened to observe a nearby galaxy with an absorption-line spectrum at a redshift of 0.542, SDSS J020748+004735, which may be in the same physical association in space.

## 4. Conclusions: Lyman Leakage from Green Pea Galaxies--Is It a Coin Flip?

The immediate goal of this observational study was to determine the incidence of escaping Lyman continuum photons from a newly discovered group of extremely luminous [OIII]-emitting galaxies at z~0.5. Having found that some but not all of them are LyC emitters, it is now interesting to compare their LyC emission with properties of other samples observed in the optical and UV which may have overlapping characteristics.

### 4.1 Difficulty of Selecting the Lyman Leakers from Optical Spectroscopy

The two green pea galaxies we observed are so similar that--without UV spectroscopy-- it is very difficult to identify any observable which could have predicted that one would be a strong leaker of Lyman continuum photons, while the other turns out not to be. They both have comparably high equivalent widths of [OIII] emission lines, comparably high specific star formation rates, comparable masses and luminosities. Both have ratios of [OIII]5007+4959/H$\beta$ > 7. J020748+004735 and J090334+174717 also have extremely high ratios of **O32 =** [OIII]5007+4959/[OII]3727--6.5 and 7.4, respectively. Thus a high ratio of O32 > 6 appears to be a necessary, but not sufficient, condition for $f_{esc,rel}$ to exceed 5%. This is consistent with the detections presented by Izotov et al. (2016a and 2018b), and also at higher redshifts by Rivera-Thorsen et al. (2019), Fletcher et al. (2019) and Vanzella et al. (2020).

Although strong [OIII] emission lines indicate sub-solar abundances, the escape of Lyman continuum photons is not simply connected to the lowest galaxy metallicities. To the contrary, the Lyman leaker we detected, J020748+004735, has log [O/H] +12 = 8.33, which, like most of our green pea sample, is still ~ one third of solar. Nor does a high $f_{esc}$ require a very low galaxy mass: the stellar mass of J020748+004735 is log $M_*$ = 9.73, significantly more than that of the LMC. Compared with previous extreme dwarf starburst galaxies which have had measured Lyman limit continuum (e.g. Izotov et al. 2016a, 2018b), our two examples are at the upper end in terms of mass and luminosity. By widening the range over which escaping Lyman limit photons have been detected, we conclude that small masses and very low metallicities are not required.

Another proposed indicator of extremely hard ionizing spectra, the detection of HeII 4686 at a level of >2% of the H$\beta$ flux (Izotov et al 2019), can be ruled out in both of these galaxies, as well as the full stack of the 22 new green pea SDSS spectra (LM18). Our measurement of the rest-frame 4686Å regions of the two individual SDSS spectra sets strict 3-sigma upper limits of HeII4686/H$\beta$ < 0.015.

Both of the target galaxies have Mg II doublet emission lines--$EW_0$(Mg II) = 13 Å and 7 Å for J020748+004735 and J090334+174717, respectively. A stronger Mg II emission line doublet has been proposed as a tentative predictor of strongly escaping Ly$\alpha$ line emission (Henry et al. 2018). Because both the MgII doublet and Ly$\alpha$ are resonantly scattered emission lines, they may require relatively low optical depths in the ISM gas to be observed strongly in emission (Feltre et al 2018). This could also make $EW_0$(Mg II) a predictor of escaping LyC photons. That hypothesis is consistent with our observations, but of course more data will be required to determine how reliable this prediction can be.

4.2 Predicting the Lyman leakers from COS observations

Although the previous observations did not give us any strong differences that allowed us to predict which of our starburst galaxies would be a detected Lyman continuum leaker, our UV observations with HST provide some possible hints:

a) Intense current star formation. The detected Lyman leaker--J020748+004735--is somewhat bluer. Its far-UV flux within the COS PSA is ~one magnitude brighter than that of J090334+174717. This is reflected in the somewhat higher specific star formation rate we infer: the SFR normalized by the stellar mass of J020748+004735 is 2.4 times higher than that of J090334+174717.

b) Compact Star formation. The detected Lyman leakers tend to have very compact UV continuum emission (e.g., Marchi et al. 2018; Izotov et al. 2018b). Both of our target galaxies are small, but the one with detectable Lyman continuum is more compact than the other one, which may be undergoing a merger.

c) Lyα emission strength. Although both of the target green pea galaxies show Lyα in emission, the higher rest frame EW(Lyα) in J020748+004735 (37Å, compared with 21Å in J090334+174717) is more comparable to those of previously detected Lyman leakers (Verhamme et al. 2017). Our intermediate values of EW(Lyα) are consistent with their observed trend that larger EW(Lyα)(greater than 50Å) is likely to be associated with LyC leakage, while smaller EW(Lyα)(less than 25Å) is not. A similar anti-correlation appears to hold at z~3 (Steidel et al. 2018, and Marchi et al. 2018).

d) Lyα emission profile. It has been suggested that the LLC leakers are more likely to have narrower central absorptions in their Lyα emission line profiles (blue/red peak separations smaller than 300--400 km/sec-- Verhamme et al. 2017; Izotov et al. 2018b). Our LLC detection of J020748+004735 supports and extends this hypothesis: it has one strong single central peak, with no apparent central absorption at all. Our non-detection of escaping LyC photons in J090334+174717 is also consistent with its observed peak separation of ~350 km/sec. Theoretical models have actually predicted that LyC-emitting galaxies should have more symmetric Lyα emission profiles, with stronger central emission peaks (Behrens, Dijkstra and Niemeyer 2014; Dijkstra, Gronke, and Venkatesan 2017), as we observe, and is also seen at high redshift (Vanzella et al. 2020).

Although Lyman continuum leakage appears to be correlated with very strong [OIII] emission and these other observables, none are perfect predictors--there is still scatter which remains unexplained. There are many possible sources of stochasticity in the ISM of these galaxies. One plausible explanation is that although the class of galaxies with extremely strong [O III] emission includes many of the best producers of escaping ionizing photons, their escape may be anisotropic, as considered by Zackrisson, Inoue and Jensen (2013) and Fletcher et al. (2019), for example. As modelled by theoretical studies (e.g. Behrens et al. 2014), the escape might occur along ionized channels, but their geometry (bipolar, "picket fence", random holes, etc.) is not known. As long as our particular viewing angle of a star-burst geometry cannot be predicted in advance, the $f_{esc}$ that we observe may include an element of luck.

Our observed sample of z~0.5 green pea galaxies has only two UV spectra, but should be representative of the entire class of luminous strong [OIII]-emitters at z=0.5 surveyed by LM18. If we did not have any UV observations, we might guess a highly uncertain ~50% success rate of detecting leaking Lyman limit continuum from these z=0.5 green pea galaxies. Nonetheless, this is too small a number on which to base a firm statistical conclusion about the probability of escaping ionizing continuum from the entire class.

### 4.3  Implications for Reionization at High Redshifts

There is growing evidence that the green pea phenomenon studied here (very strong [OIII] line emission) was more widespread in galaxies at higher redshifts. It is extremely rare in the low-z universe (Cardamone et al. 2009), but the ``knee'' in the [OIII] line luminosity function brightens by 2.5 times from z=0.41 to z=0.83 (Ly et al. 2007).  And the evolution continues strongly to higher redshifts.  At z~1.5, ultra-strong [OIII] emitters are ~10 times more numerous than at z~0.1 (Maseda et al. 2018). Similarly, Atek et al. (2011) used Hubble Space Telescope infrared grism spectroscopy to discover ~a hundred [OIII] emitting galaxies at z~2 (with rest equivalent widths from 200 to 400A), in only 180 square arcminutes. Atek et al. (2014) then quadrupled this galaxy sample, and demonstrated that the overwhelming majority of them are extreme starbursts.  By z=3.5, stacked spectral energy distributions (SEDs) of thousands of U-dropout galaxies show that as the stellar mass of the *average* Lyman break galaxy drops from $10^9$ to $10^8$ Suns, the rest EW([OIII]5007+4959) rises from 400 to >2000Å, showing that 'green peas' are ubiquitous at z>3 (Malkan et al. 2017). Such high-ionization starbursts are evidently even more universal at z>6, as indicated by their IRAC-band excess flux due to extremely strong [OIII] emission in the broad-band infrared filters (Smit et al. 2015). Thus, an extrapolation of the results in this work and others to high redshifts suggests that they are the best candidates for re-ionizing the universe.

## Acknowledgments

We thank Dr. Alaina Henry for a valuable critical reading of a draft of this work, and the referee, for a very helpful detailed review, which greatly improved the paper. This research work was supported by NASA/STSci Guest Observer Grant GO-15471. It has made use of the data obtained by SDSS and WISE. Funding for the Sloan Digital Sky Survey IV has been provided by the Alfred P. Sloan Foundation, the U.S. Department of Energy Office of Science, and the Participating Institutions. SDSS-IV acknowledges support and resources from the Center for High-Performance Computing at the University of Utah. SDSS-IV is managed by the Astrophysical Research Consortium for the Participating Institutions of the SDSS Collaboration including the Brazilian Participation Group, the Carnegie Institution for Science, Carnegie Mellon University, the Chilean Participation Group, the French Participation Group, Harvard-Smithsonian Center for Astrophysics, Instituto de Astrofısica de Canarias, The Johns Hopkins University, Kavli Institute for the Physics and Mathematics of the Universe / University of Tokyo, Lawrence Berkeley National Laboratory, Leibniz Institut fur Astrophysik Potsdam, Max-Planck-Institut fur Astronomie, Max-Planck-Institut fur Astrophysik, Max-Planck-Institut fur Extraterrestrische Physik, National Astronomical Observatories of China, New Mexico State University, New York University, University of Notre Dame, Observatario Nacional/MCTI, The Ohio State University, Pennsylvania State University, Shanghai Astronomical Observatory, United Kingdom Participation Group, Universidad Nacional Autonoma de Mexico, University of Arizona, University of Colorado Boulder, University of Oxford, University of Portsmouth, University of Utah, University of Virginia, University of Washington, University of Wisconsin, Vanderbilt University, and Yale University. WISE is a joint project of the University of California, Los Angeles, and the Jet Propulsion Laboratory of the California Institute of Technology, funded by the National Aeronautics and Space Administration. The WISE web site is http://wise.astro.ucla.edu/.

Facilities: Hubble Space Telescope, Lick Observatory, Sloan Digital Sky Survey

Software: IRAF, Python

Orcid ID: Matthew A. Malkan https://orcid.org/0000-0001-6919-1237